\documentclass[useAMS,usenatbib]{mn2e}
\usepackage{epsfig}
\usepackage{times}
\usepackage{amssymb}

\title[Millennium Simulation: halo formation times]{A marked correlation
  function analysis of halo formation times in the Millennium
  Simulation}

\author[G. Harker et al.]{Geraint Harker,\thanks{E-mail:
  g.j.a.harker@durham.ac.uk} Shaun Cole, John Helly,
  Carlos Frenk and Adrian Jenkins\\
  Department of Physics, University of Durham, Science Laboratories,
  South Road, Durham DH1 3LE}

\begin{document}

\date{\today}

\maketitle

\begin{abstract}
We study the environmental dependence of the formation epoch of dark
matter haloes in the Millennium Simulation: a ten billion particle
$N$-body simulation of standard $\Lambda$CDM cosmology. A sensitive
test of this dependence -- the marked correlation function -- reveals
highly significant evidence that haloes of a given mass form earlier
in denser regions. We define a marked cross-correlation function,
which helps quantify how this effect depends upon the choice of the
halo population used to define the environment. The mean halo
formation redshift as a function of the local overdensity in dark
matter is also well determined, and we see an especially clear
dependence for galaxy-sized haloes. This contradicts one of the basic
predictions of the excursion set model of structure formation, even
though we see that this theory predicts other features of the
distribution of halo formation epochs rather well. It also invalidates
an assumption usually employed in the popular halo, or HOD, models of
galaxy clustering, namely that the distribution of halo properties is
a function of halo mass but not of halo environment.
\end{abstract}

\begin{keywords}
cosmology: theory -- dark matter -- galaxies: haloes -- galaxies: formation
\end{keywords}

\section{Introduction}

In the Cold Dark Matter paradigm, the large-scale structure of the
Universe results from the amplification and evolution under gravity
of small initial perturbations in the density distribution of massive,
collisionless particles. On smaller scales, the clustering becomes
nonlinear, and the dark matter collapses into relatively dense,
virialized clumps -- dark matter haloes. Gas falls into the potential
wells created by these haloes, where it can cool and form stars
\citep{WHI78}. It then seems as though the problem of understanding
the properties and clustering of galaxies splits naturally into two
parts -- understanding the distribution of the haloes, and
understanding the processes involving the dark matter and the baryonic
components inside the haloes. This is clearly simplified if the two
parts of the problem can be considered independently.

In current semi-analytic models of galaxy formation, the evolution of
galaxies in haloes is driven by the merger histories of those haloes
\citep[e.g.,][]{KAU93,COL94,SOM99,COL00}. We expect to be able to treat
haloes independently of the large-scale structure if their merger
histories are independent of the large-scale environment. This is also
the basis of the halo occupation distribution (HOD) formalism
\citep{SEL00,BER02,COO02}. Support for making this simplification has
come from extended Press-Schechter theory \citep{BCEK,BOW91,LAC93},
which predicts that the distribution of halo formation times is a
function of halo mass but not of halo environment. This theory is the
basis for the construction of Monte Carlo merger trees for
semi-analytic models. To reach this conclusion, however, three
simplifying assumptions are made, which if relaxed may result in an
environmental dependence. Firstly, in order to solve the
cloud-in-cloud problem \citep{BBKS} it is assumed that the
trajectories, $\delta(M)$, in overdensity versus mass scale can be
treated as Brownian random walks. This is only true when the density
field is filtered with a sharp $k$-space filter. For a more natural
top-hat or Gaussian filter the $\delta(M)$ trajectories exhibit
correlations between different mass scales, which induce correlations
between environment and small scale behaviour
\citep{BCEK,LAC93}. Secondly, the extended Press-Schechter or
excursion set theory deals only with individual mass points in the
density field. When it predicts that a mass point is part of a halo of
mass $M$ there is no constraint that the whole of a neighbouring
volume of mass $M$ is also assigned to the same halo. The accuracy to
which this assumption holds may depend on environment. Thirdly, in
determining when a region collapses to form a halo, a global collapse
threshold given by the spherical collapse model is assumed. It has
been argued that tidal fields modify this threshold \citep{SMT01} and
these could also depend on environment.

Thus there is no compelling reason to believe that the lack of
environmental dependence predicted by the Press-Schechter theory
should carry over to a full treatment which relaxes these
assumptions. It has been supported by $N$-body calculations, however,
e.g.\ by \citet{LEM99} who used the GIF simulations
\citep{JEN98,KAU99}. Provided such simulations have sufficiently many
outputs at different times, merger trees can be extracted and their
environmental dependence studied. This approach has been limited by
the dynamic range of the available simulations. Either galaxy-sized
haloes have not been well resolved, leaving their merger histories
uncertain, or the volume has not been cosmologically representative.

Galaxy properties, however, do depend on environment. Galaxies in
denser regions tend, for example, to be more bulge-dominated and to
have older stellar populations
\citep[e.g.,][]{DRE80,POS84,GOM03,BAL04}. In models in which the
merger histories of haloes are only a function of halo mass, and in
the absence of non-local gas processes (e.g.\ ionization by QSOs),
this can only be accounted for by the variation of the halo mass
function with environment, or, in other words, by the fact that high
mass haloes are more clustered than low mass haloes
\citep{COL89,MO99}. Models which attempt to reproduce environmental
dependence can then do so only by populating more massive haloes with
a greater fraction of early-type galaxies.

None the less, \citet{SHE04b} argue that one of the results of
\citet{LEM99} suggests, rather indirectly, an environmental dependence
of halo formation times. We revisit this argument in
Section~\ref{sec:results}, noting that it also predicts the sign of
the dependence, and predicts it correctly in the light of our
results. The range of assumptions required for analytic theory to
predict environmental independence also suggests that detection of
some signal should be possible. To make progress on this matter using
$N$-body simulations seems, then, to require one of two
things. Firstly, we may try to pin down the environmental dependence
of halo formation times suggested by the above results by using more
sensitive tests. \citet{SHE04b} claimed to have found such a test --
the marked correlation function -- and found a signal of environmental
dependence despite using the same GIF simulations as
\citet{LEM99}. Marked statistics have recently proved useful in the
analysis of both simulations \citep{FAL02,GOT02,SHE06} and surveys
\citep{BEI00}, offering both sensitivity and information complementary
to that provided by other statistics. A more general discussion of
marked statistics and their interpretation may be found in
\citet{SHE05}. Secondly, we may use larger simulations, so that even a
subset of the haloes spanning a small range in mass provides adequate
statistical power to see significant evidence of environmental
dependence, if this dependence exists and is sufficiently large to be
interesting. Higher resolution would also allow us to study haloes
which host only a single bright galaxy, so that we may hope for a more
direct link between the halo properties and the galaxy properties than
one would expect when studying more massive haloes. An environmental
dependence of the merger histories of galaxy-sized haloes may provide
a more direct explanation for the variation in galaxy properties with
environment, and would suggest that the systematic change in the halo
mass function with environment is not the only driving force behind
the systematic change in galaxy properties with environment.

In this paper we attempt to combine both the above techniques. That
is, we calculate the marked correlation function as suggested by
\citet{SHE04b}, and later go on to discuss some other statistics
closely related to the marked correlation function. We apply these
calculations to the ``Millennium Simulation'' \citep{SPR05b}, which
resolves the merger histories of haloes small enough that we expect
them to host a single galaxy of luminosity $0.1L_\ast$ (where $L_\ast$
is the characteristic luminosity corresponding to the break in the
galaxy luminosity function), but that probes a cosmologically
representative volume. This is the simulation used by {Gao},
{Springel} \& {White} (2005)\nocite{GAO05} to study the age-dependence
of halo clustering, using an approach which is complementary to that
taken here.

The structure of the paper is as follows. In Section~\ref{sec:milsim}
we describe the Millennium Simulation, and the merger trees used to
calculate the formation times in this work. We describe the marked
correlation function in Section~\ref{sec:ximark}. We also discuss here
the choice of mark used for the majority of the results presented in
this paper. Then in Section~\ref{sec:results} we go on to describe our
results, including tests to justify our choice of mark. These motivate
the definition of a marked cross-correlation function, which we
calculate for various halo samples. We also present results of a test
of the effect of environment on halo formation time which corresponds
more directly to earlier calculations using smaller
simulations. Finally, we present our conclusions in
Section~\ref{sec:conc}.

Throughout we use the convention that the Hubble constant, \hbox{$H=100h\ \mathrm{km\ s^{-1}\ Mpc^{-1}}$}.

\section{The Simulation}\label{sec:milsim}

In this study we use the Millennium Simulation \citep{SPR05b} carried
out by the Virgo Consortium using a modified version of the {\small
TREE-PM} $N$-body code {\small GADGET2} \citep{SPR01b,SPR05a}. The
cosmology is a flat, $\Lambda$CDM model, with $\Omega_\mathrm{m}=0.25$
(so $\Omega_\Lambda=0.75$) and $h=0.73$. The initial power spectrum
was calculated using {\small CMBFAST} \citep{SEL96}, and is such that
the primordial power spectrum has power-law index $n=1$, the {\it rms}
linear mass fluctuation in spheres of radius $8\ h^{-1}\ \mathrm{Mpc}$
extrapolated to $z=0$ is $\sigma_8=0.9$, and the baryon density is
$\Omega_\mathrm{b}=0.045$. This leaves a dark matter density,
$\Omega_\mathrm{dm}=0.205$. The simulation follows the evolution under
gravity of $2160^3$ dark matter particles in a periodic box with sides
of comoving length $500\ h^{-1}\ \mathrm{Mpc}$ from $z=127$ to the
present day. Each particle has mass $8.61\times 10^8\ h^{-1}\
\mathrm{M}_\odot$, and the gravitational force has a
Plummer-equivalent comoving softening length of $5\ h^{-1}\
\mathrm{kpc}$. The particle data were output and stored at 64 times,
60 of which are spaced regularly in the logarithm of the
expansion factor between $z=20$ and $z=0$, allowing the construction
of trees detailing how each dark matter halo at $z=0$ was built up
through mergers and accretion.

\subsection{Merger trees}\label{subsec:mtrees}

At each of the output times of the simulation we have a catalogue of
friends-of-friends groups \citep{DAV85} calculated using a linking
length of $b=0.2$ times the mean inter-particle separation. Locally
overdense, self-bound substructures of these groups are found using
the {\small SUBFIND} algorithm \citep{SPR01a}. Each friends-of-friends
halo is therefore decomposed into a collection of subhaloes, plus a
fuzz of unbound particles. Of the subhaloes, one is typically much
larger than the others and contains most of the mass of the halo. This
can be thought of as the background mass distribution of the halo,
while the smaller subhaloes are substructures.

Sometimes, however, the friends-of-friends algorithm links together
structures which one might prefer to consider as separate haloes for
the purpose of constructing the merger trees. Visually, these haloes
often appear to consist of two distinct structures joined by a tenuous
bridge of particles. They may also be only temporarily joined, in the
sense that following the evolution of the system would see the
structures move apart and become distinct friends-of-friends haloes
again. Having run {\small SUBFIND}, we identify these spuriously
linked haloes as follows. We split a subgroup from its
friends-of-friends halo before calculating the merger trees if either
of the following conditions is satisfied: the centre of the subhalo is
outside twice the half-mass radius of the main halo; or the subhalo
has retained more than 75 per cent of the mass it had at the last
output time at which it was an independent halo. The latter condition
is imposed because we expect a less massive halo to be stripped of its
outer layers as it merges with a more massive halo, while if it has
been artificially linked by the friends-of-friends algorithm it will
have retained most of its mass. Treating the subgroups which have been
split off as separate haloes has also been found to lead to a better
match between galaxy properties in SPH simulations and in semianalytic
models which use the resulting merger trees \citep{HEL03b}.

The splitting algorithm above results in a halo catalogue containing
more haloes than in the original friends-of-friends catalogue.  When
we refer to a `halo' below, we refer to a member of this new, larger
catalogue.  A halo, as before, is a collection of {\small SUBFIND}
subhaloes including one background subhalo. Each halo in the catalogue
at the final time has its own merger tree built from these catalogues.
It has become conventional in studying the properties of the merger
trees themselves, however, to calculate one merger tree per
friends-of-friends halo, i.e.\ to define a halo as a
friends-of-friends object. To provide contact with earlier work,
therefore, if the splitting algorithm above results in a
friends-of-friends halo being associated with two or more `merger tree
haloes' at the final time, we consider only the merger tree of the
most massive component, and discard the other trees from the same halo
in our analysis.  The merger tree of this remaining component is
unaffected by discarding the less massive components, since each
subhalo at each redshift may appear in only one merger tree (in other
words, if a halo or subhalo at some time has a descendant at a later
time, as almost all haloes do, then this descendant is
unique). Approximately 15 per cent of haloes are split in this way,
and usually the mass of the discarded part is only a small fraction of
the mass of the halo. The proportion of split haloes decreases with
increasing halo mass, dropping to only a few per cent for haloes with
mass close to the characteristic mass, $M_\ast$.

The merger trees are constructed from the group catalogues by
following subhaloes from early times to late times, identifying in
which halo a subhalo resides at the later time (Helly et al., in
prep.). This means that given a subhalo in one snapshot, we must be
able to find the corresponding object (the descendant subhalo) in a
later snapshot. This is usually the next snapshot, though we check for
a descendant in the next five outputs since occasionally
friends-of-friends or {\small SUBFIND} are unable to identify the
subhalo in the intervening snapshots. This may happen when, for
example, a halo loses particles and drops below the resolution limit,
or passes through a dense region in which it is not identified as a
distinct object. The descendant of a subhalo is found by following the
most bound 10 per cent of its mass or the 10 most bound particles,
whichever is the greater mass. The descendant is the subhalo which
contains the largest number of these particles. We identify the
descendant of an entire halo as being the halo which contains the
descendant of its most massive subhalo. Haloes therefore do not split:
a halo at redshift $z_1$ has at most one descendant at redshift
$z_2<z_1$. If the particles of a halo do become distributed between
two haloes at a later time, only one of these two haloes may have the
original halo as a progenitor. De-merger events may therefore lead to
`orphan' haloes with no progenitors. This physical splitting or
de-merger of haloes as the simulation evolves is unrelated to the
algorithm we use to split friends-of-friends haloes above. Clearly,
though, our definition of a halo affects whether or not we consider
two haloes to have de-merged, and we comment briefly on the impact of
de-mergers on our results in Section~\ref{subsec:oldtest}.

Given a parent halo in the final snapshot, we call all haloes in
earlier snapshots whose descendants are within the halo its
progenitors. At each of the earlier snapshots, one of the progenitor
haloes is designated the `main' progenitor of the parent halo. This
main progenitor is defined inductively as we move up in redshift one
snapshot at a time as the most massive progenitor of the main
progenitor in the previous snapshot. We then define the formation time
of a halo as the redshift at which the main progenitor had half the
mass of the final halo, linearly interpolating between the two
redshifts at which its mass was greater than and less than half the
final mass.  This definition of formation redshift -- the redshift at
which the mass of the main progenitor falls below half the mass of the
final halo -- provides contact with analytic approaches to this
problem and with earlier work on the formation time of $N$-body haloes
\citep{LAC93,SHE04b}.

To calculate a marked correlation function of haloes we need to know
the distance between any two haloes. We define this as the distance
between their centres, and take the centre to be the position of the
particle with the minimum gravitional potential energy, which is
output by {\small SUBFIND}.

Finally, note that the trees used in this work were constructed
independently of the Millennium Simulation merger trees discussed by
\citet{SPR05b} and \citet{GAO05}. The two sets of trees differ both in
the criteria for identifying independent haloes and in the treatment
and identification of the descendant haloes themselves. In this
respect, and in respect of the methods we use to analyse our halo
catalogues, this work complements the study of the environmental
dependence of halo formation by \citet{GAO05}. A discussion of the
issues involved in constructing suitable merger trees (especially in
the context of semianalytic models of galaxy formation) may be found
in \citet{HEL03a}.

\section{The marked correlation function}\label{sec:ximark}

Studying the dependence of halo formation time on halo environment
requires, of course, a definition of halo environment. When using a
dark matter simulation, a natural definition is the local overdensity
in dark matter, measured on some chosen scale. This immediately
highlights the problem of choosing an appropriate scale. It is not
clear, for example, whether the choice of scale should depend on the
mass of the halo under consideration. Then there are subsidiary
choices such as whether to excise the region containing most of the
mass of the halo from the region used to define the local
overdensity.

\citet{LEM99} studied halo formation time as a function of the
overdensity of dark matter in a spherical shell of inner radius
\hbox{$2\ h^{-1}\ \mathrm{Mpc}$} and outer radius \hbox{$5\ h^{-1}\
\mathrm{Mpc}$} centred on the halo. There was no significant detection
of a dependence of formation time on environment defined in this
way. \citet{SHE04b}, however, proposed a test which they considered
more sensitive, and which does not require a similar choice of
scale. Their `marked correlation function' is defined as follows.

Consider a set of $N$ objects, taken in this case to be dark matter
haloes.  To each one assign a `mark' \hbox{$\{m_i\ ,\ i=1,\ldots ,N\}$},
where in this study we take the mark to be formation redshift, or some
proxy for formation redshift. Let the pair \hbox{$\{i,j\}$} have separation
\hbox{$r_{ij}$}. Then the marked correlation function
\hbox{$\xi_\mathrm{marked}(r)$}, a function of separation $r$, is defined by
\begin{equation}
\xi_\mathrm{marked}(r)=\sum_{\{i,j\ |\ r_{ij}=r\}}\frac{m_i m_j}{n(r)\bar{m}^2}\quad ,
\end{equation}
where \hbox{$n(r)$} is the number of pairs of objects with separation
\hbox{$r_{ij}=r$} and where the mean mark $\bar{m}$ is calculated over
all objects (of all separations) in the sample.

In other words, if \hbox{$\xi_\mathrm{marked}(r)>1$} for some $r$ then
this implies that pairs of objects with separation $r$ have a greater
value of the mark than average.  In the case of dark matter haloes, we
expect that haloes in overdense environments have more close
neighbours than haloes in underdense environments (some caveats to
this interpretation are discussed in
Section~\ref{sec:results}). Therefore the contribution of haloes in
overdense environments dominates \hbox{$\xi_\mathrm{marked}(r)$} on
small scales. On large scales, meanwhile, we expect to recover the
global average, \hbox{$\xi_\mathrm{marked}(r)=1$}. If we see that
\hbox{$\xi_\mathrm{marked}(r)$} deviates from $1$ on some scale we may
interpret this as an environmental dependence of the mark.

Note we do not have to choose a scale on which to study this
dependence; the marked correlation function tells us the scale. This
is clearly desirable, but comes at the cost that there is no
straightforward correspondence between environment as defined by the
marked correlation function and environment as defined by the
overdensity in some region near the halo.

\subsection{Choice of mark}\label{subsec:mchoice}

In principle one could choose to measure the marked correlation
function using any of a whole range of halo properties as the mark, in
order to investigate the environmental dependence of those properties.
Here, although we wish to study the environmental dependence of halo
formation redshift, it may not be best to use this as the
mark. Instead, for the majority of our results we follow
\citet{SHE04b} and use a `scaled formation redshift' for our mark. The
definition of scaled formation redshift, used here and in
\citet{SHE04b}, is the $\tilde{\omega}_\mathrm{f}$ parameter defined
in equation~2.31 of \citet{LAC93}. Suppose we measure formation
redshift relative to some final time $z_0$ (here, we always take $z_0=0$),
and consider a halo with mass $M_0$ at $z_0$ and which formed at a
redshift $z_\mathrm{f}$. Then $\tilde{\omega}_\mathrm{f}$ is given by
\begin{equation}
\tilde{\omega}_\mathrm{f}=\frac{\delta_\mathrm{c}(z_\mathrm{f})-\delta_\mathrm{c}(z_0)}{\sqrt{\sigma^2(M_0/2)-\sigma^2(M_0)}}\quad
,\label{eq:omtildedef}
\end{equation}
where \hbox{$\delta_\mathrm{c}(z)$} is the critical density threshold for
collapse and \hbox{$\sigma^2(M)$} is the linear theory variance of density
fluctuations at mass scale $M$.

\begin{figure}
  \begin{center}
    \leavevmode
    \psfig{file=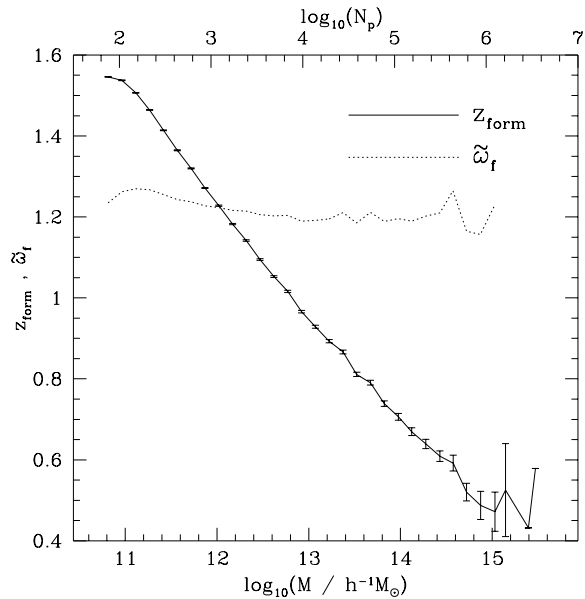,width=8cm}
    \caption{The solid line with error bars gives the mean formation
    redshift of haloes in the Millennium Simulation as a function of
    the mass of the halo (lower horizontal axis) or, equivalently, the
    number of particles in the halo (upper horizontal axis).  The
    dotted line, which exhibits a much weaker mass dependence, shows
    on the same scale the mean value of the scaled formation redshift,
    $\tilde{\omega}_\mathrm{f}$ (see Equation~\ref{eq:omtildedef} for
    a definition), of haloes as a function of halo
    mass.}\label{fig:zvsnp}
  \end{center}
\end{figure}

The motivation for using $\tilde{\omega}_\mathrm{f}$ rather than
$z_\mathrm{f}$ as the mark comes from the following predictions of
extended Press-Schechter theory: firstly, that the distribution of
$\tilde{\omega}_\mathrm{f}$ depends very weakly on the initial power
spectrum of fluctuations; and secondly, that for a power-law initial
power spectrum, the distribution of $\tilde{\omega}_\mathrm{f}$ is
independent of halo mass. The latter prediction still holds to very
high accuracy for more general power spectra with slowly varying
slope.  Moreover, the prediction is largely confirmed by measurement
of the distribution in our simulation. This is demonstrated in
Fig.~\ref{fig:zvsnp}, where the mean formation redshift of haloes in
the Millennium simulation is plotted as a function of mass. For
comparison, we plot the mean value of $\tilde{\omega}_\mathrm{f}$ as a
function of mass on the same scale. Clearly
$\tilde{\omega}_\mathrm{f}$ scales out much of the dependence of halo
formation redshift on halo mass.

\begin{figure}
  \begin{center}
    \leavevmode
    \psfig{file=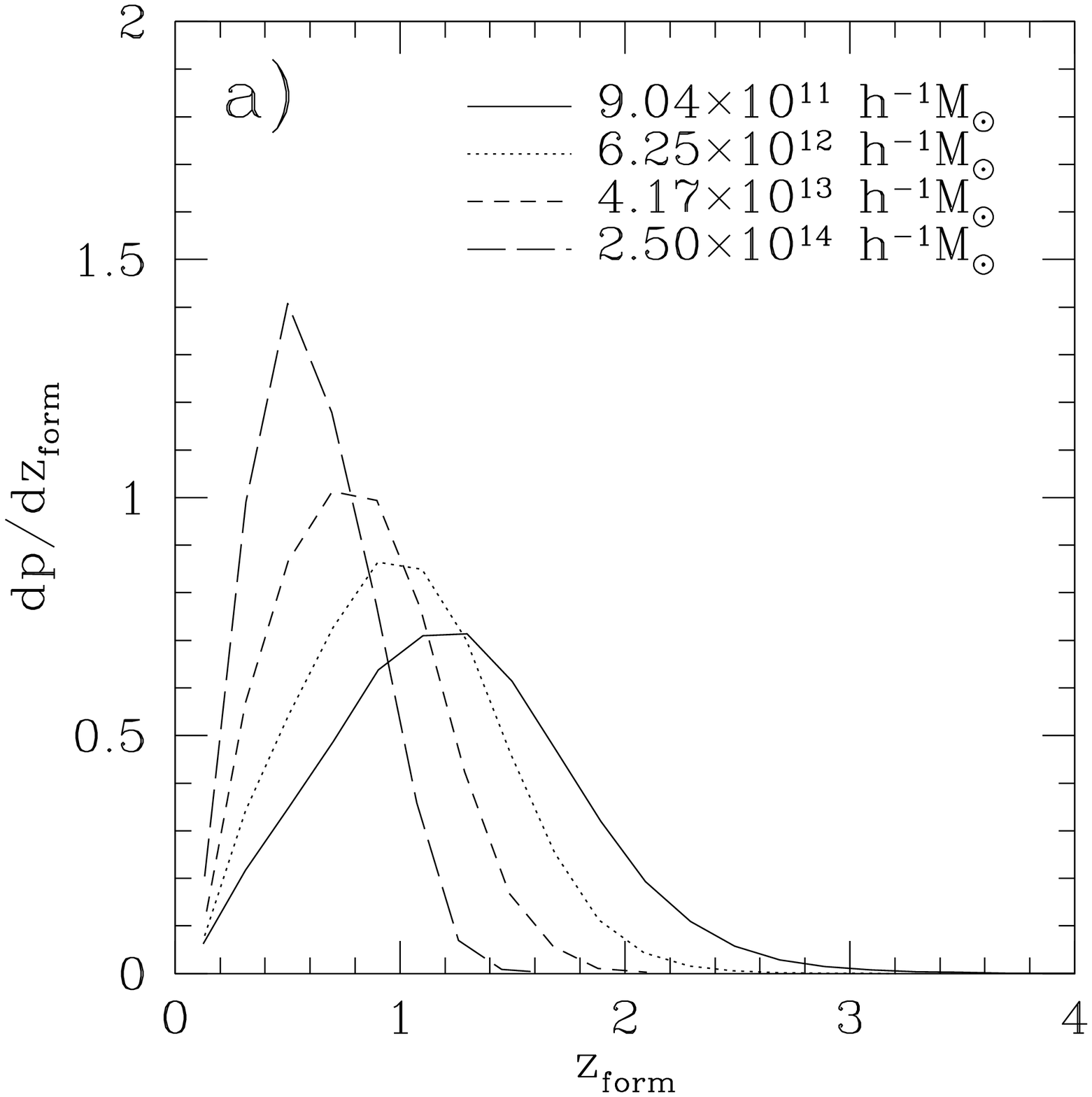,width=8cm}
%
    \psfig{file=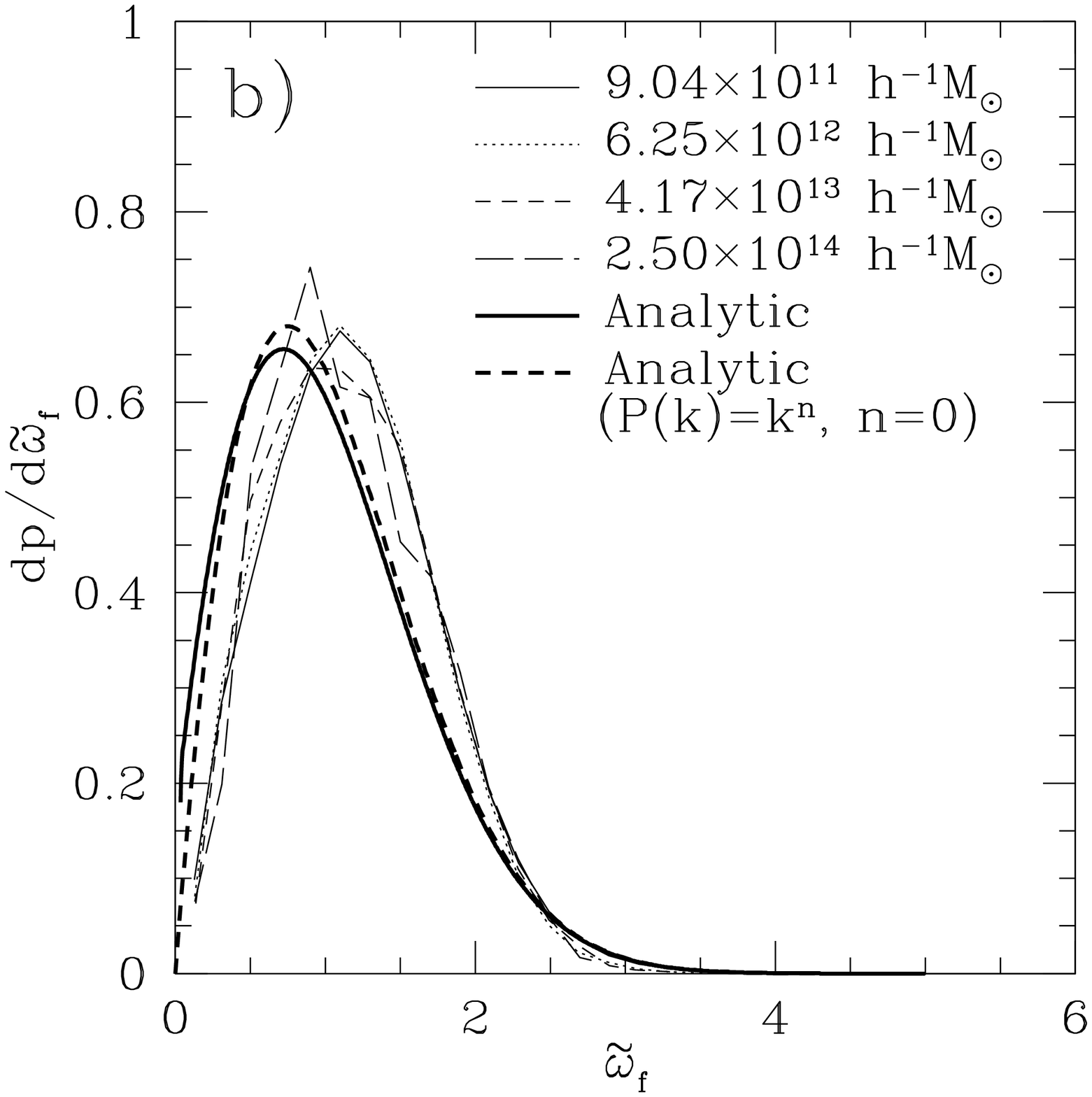,width=8cm}
    \caption{The distribution of halo formation redshift (top panel)
    and of scaled formation redshift $\tilde{\omega}_\mathrm{f}$
    (bottom panel). The distribution is shown for haloes in four
    different mass bins spaced equally in $\log(\mathrm{halo\ mass})$
    and centred on the mass given in the legend. The trend in the top
    panel is such that haloes of larger mass have a more strongly
    peaked distribution, with the peak at smaller redshift. In the
    bottom panel, we show in addition the analytic prediction for the
    distribution of $\tilde{\omega}_\mathrm{f}$ (which is very nearly
    independent of mass) with thicker, smoother lines. The thick,
    solid line shows the prediction using the input power spectrum for
    the Millennium Simulation, while the thick, dashed line shows the
    prediction using a power-law initial power spectrum with index
    $n=0$.}\label{fig:dpdzw}
  \end{center}
\end{figure}

This can be seen in more detail by comparing Fig.~\ref{fig:dpdzw}a,
which shows the distribution of formation redshift for haloes in
different mass bins in the simulation, with Fig.~\ref{fig:dpdzw}b, which
shows the distribution of $\tilde{\omega}_\mathrm{f}$ for the
corresponding haloes. In Fig.~\ref{fig:dpdzw}a we can easily see that
haloes of different masses have very different distributions of
formation redshift, and that there is a clear trend of larger mass
haloes having a more strongly peaked distribution with a peak at
smaller redshift.  In Fig.~\ref{fig:dpdzw}b, however, we see that the
distribution of $\tilde{\omega}_\mathrm{f}$ is quite similar for
haloes of different mass, and that there is no such clear trend.
Fig.~\ref{fig:dpdzw}b also shows the analytic prediction for this
distribution, which can be seen to be a reasonable approximation. The
analytic form captures the shape of the distribution well, though it
appears to predict a distribution peaking at smaller
$\tilde{\omega}_\mathrm{f}$. We show both the analytic distribution
calculated using the actual input power spectrum of the Millennium
Simulation, and the closed form for a power-law initial power spectrum
with index \hbox{$n=0$} \citep{LAC93}. Note the very weak dependence
of the distribution on power spectrum.

The main benefit of defining our mark in this way is that we may now
be justified in calculating the marked correlation function for a set
of haloes which span a broad range in mass, thereby utilising the full
statistical power of our simulation. Such a function would not have
been easy to interpret using $z_\mathrm{f}$ as the mark, since it is
well established that the halo mass function depends on local density:
in high-density regions, it becomes skewed towards more massive haloes
\citep{FRE88,COL89,LEM99,GOT03,MO04}. Because these more massive haloes tend to
have formed more recently, we could not have been sure that any signal
in the marked correlation function was not due merely to the
environmental dependence of the mass function, rather than of mean
halo formation redshift for haloes of a given mass. This effect could
also have swamped any genuine signal from an environmental dependence
of formation time.

\section{Results and extensions}\label{sec:results}

\begin{figure}
  \begin{center}
    \leavevmode \psfig{file=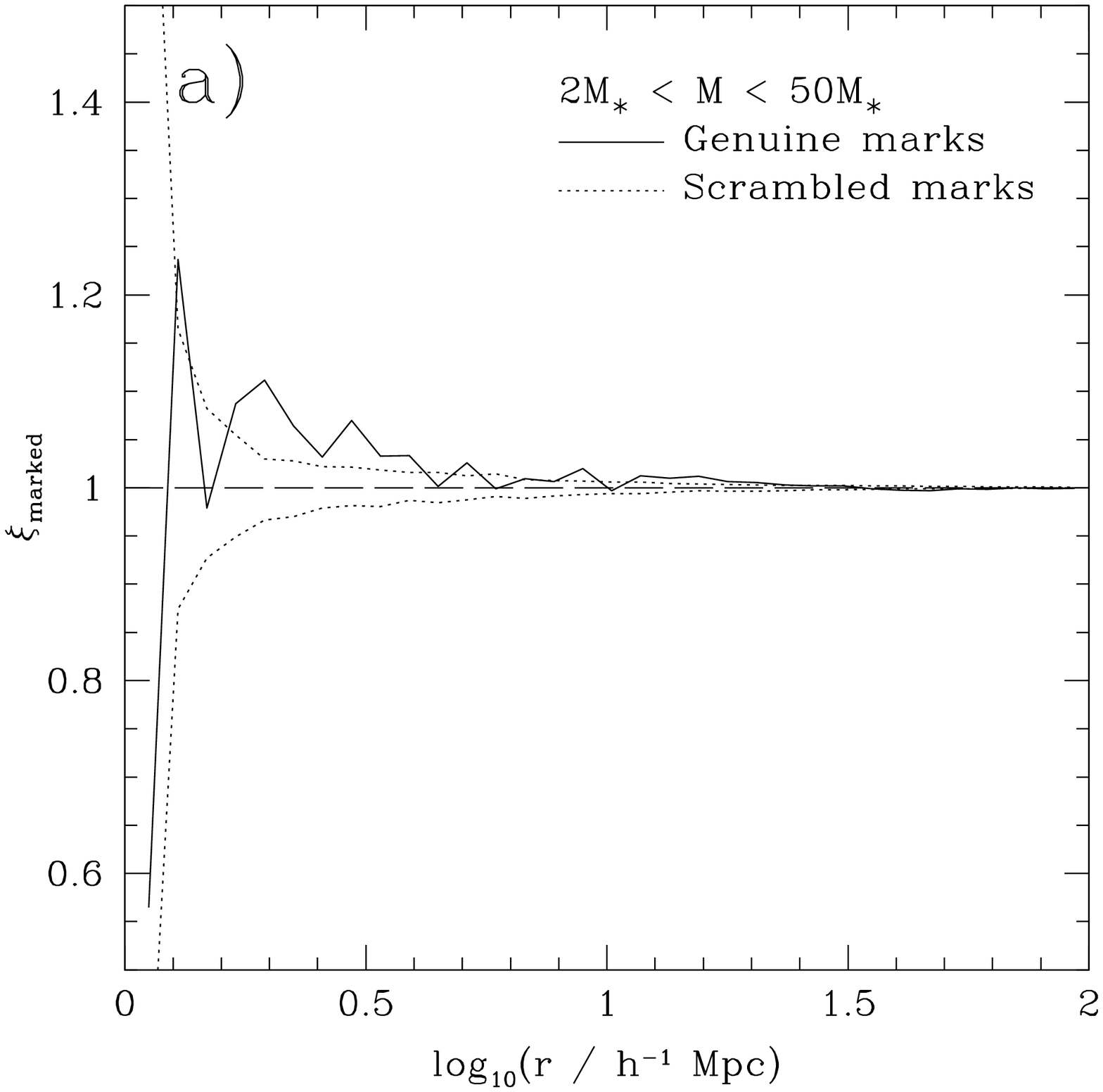,width=8cm}
    \psfig{file=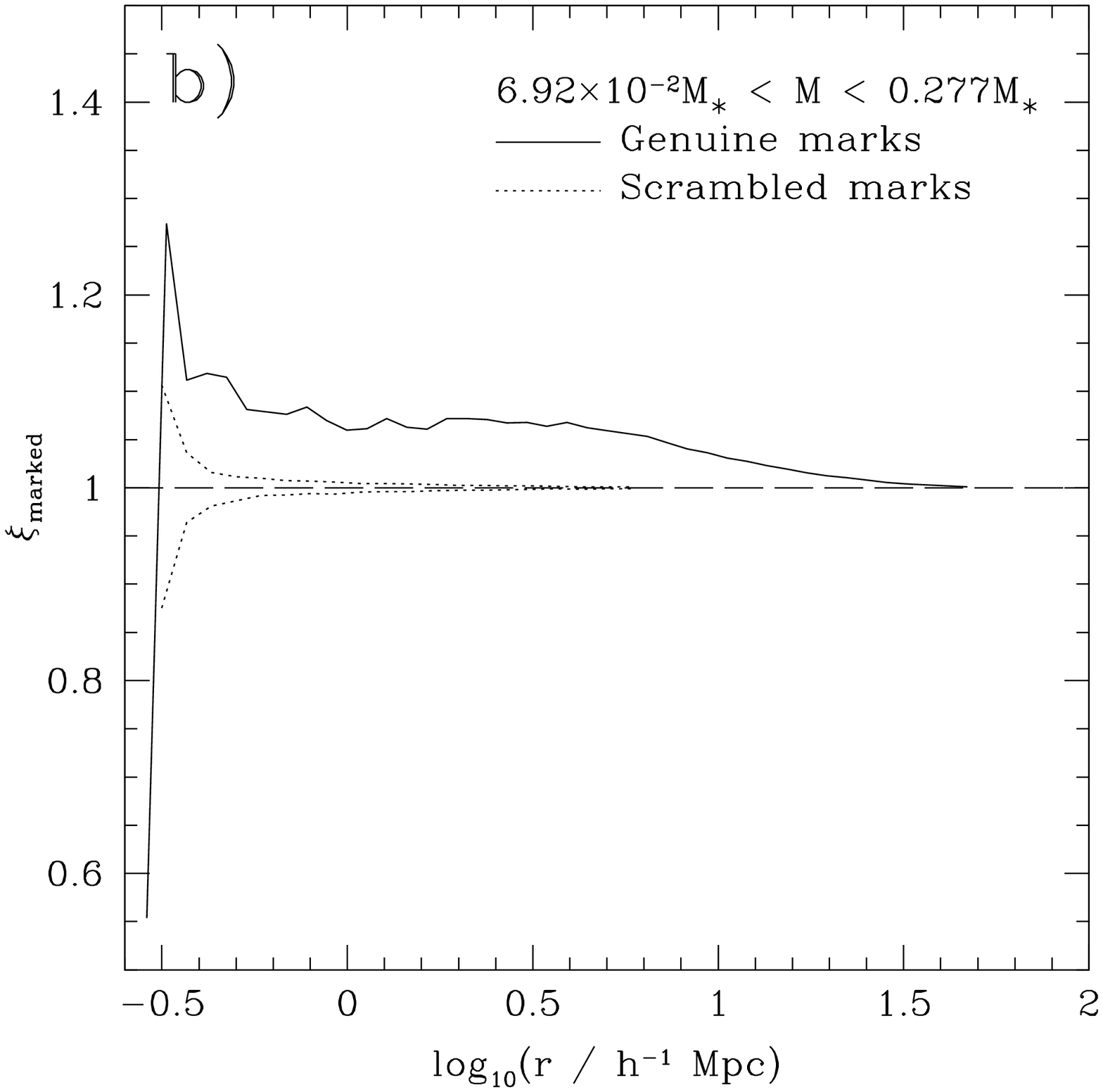,width=8cm}
    \caption{The marked correlation function (using
    $\tilde{\omega}_\mathrm{f}$ as the mark) of haloes with mass, $M$,
    in the range shown (solid line).  In the top panel, this
    corresponds to haloes with a number of particles,
    \hbox{$N_\mathrm{p}$}, such that \hbox{$14\,441\leq
    N_\mathrm{p}\leq 361\,036$} (there are 7221 particles in an
    $M_\ast$ halo). There are 34\,241 haloes in this mass range in the
    Millennium Simulation. In the bottom panel, the haloes have
    between 500 and 2000 particles. These haloes have a mass such that
    we typically expect them to host a single bright galaxy.}
    \label{fig:onehilomass}
  \end{center}
\end{figure}

A calculation of the marked correlation function of haloes with mass
between \hbox{$3.11\times 10^{12}\ h^{-1}\ \mathrm{M}_\odot$} and
\hbox{$3.11\times 10^{14}\ h^{-1}\ \mathrm{M}_\odot$} at \hbox{$z=0$}
is given in Fig.~\ref{fig:onehilomass}a (our results will be for
\hbox{$z=0$} throughout). We write the halo mass in terms of the
characteristic mass $M_\ast$, where $M_\ast$ is defined in the usual
way such that \hbox{$\sigma(M_\ast)=\delta_\mathrm{c}$}, and where
\hbox{$\delta_\mathrm{c}(z=0)=1.674$} for the cosmology assumed
here. $M_\ast$ haloes are both well resolved and numerous, containing
7221 particles and having a mass of \hbox{$6.21\times 10^{12}\ h^{-1}\
\mathrm{M}_\odot$} at \hbox{$z=0$} in the Millennium Simulation. The
peak in the function at intermediate scales indicates that haloes in
pairs with these separations have a mean formation redshift which is
higher than the global average for haloes of this mass. The function
tends to 1 at large scales, as expected. At smaller scales than those
plotted, i.e.\ less than approximately \hbox{$1\ h^{-1}\
\mathrm{Mpc}$}, the marked correlation function is not defined for
haloes of this mass, since there are no pairs of haloes in this mass
range at such small separations. Clumps of mass closer than this will
tend to be identified as part of the same structure by the
group-finder.

The sense of the dependence (higher formation redshifts in denser
regions) is that predicted by \citet{SHE04b} from the results of
\citet{LEM99}. They noted that when the distribution of formation
times (averaged over haloes of all mass) was plotted for haloes
residing in regions of different overdensity (measured in a spherical
shell between $2$ and \hbox{$5\ h^{-1}\ \mathrm{Mpc}$} centred on the
halo), the curves were very similar, i.e.\ the distribution of halo
formation redshifts was independent of local density. This seems
inconsistent with the fact that denser regions tend to host more
massive haloes, which have, on average, more recent formation times
(see Fig.~\ref{fig:zvsnp}). One might expect that because the
distribution is calculated by averaging over all haloes for each bin
in overdensity, the distribution should shift to lower formation
redshifts in more dense regions, but this was not observed. This could
be explained if haloes of a given mass tend to have higher formation
redshifts in more dense regions. No such signal was observed in the
GIF simulations, which motivates the use of a more sensitive test of
environmental dependence. It also suggests using simulations of larger
volume, since while the volume of the GIF simulations may have been
sufficient to detect a variation in the distribution of formation
times when averaging over haloes of all masses, it was not sufficient
for \citet{LEM99} to detect a variation in the mean formation redshift
as a function of local overdensity for haloes in some narrow range in
mass. The Millennium Simulation offers the opportunity to do this (and
to extend the study to haloes of lower mass) and we do so in
Section~\ref{subsec:oldtest}.

Rather than plot error bars on the (correlated) points of
Fig.~\ref{fig:onehilomass}a, we attempt to assess the significance of any
signal similarly to \citet{SHE04b}. That is, we take the population of
haloes used to calculate the marked correlation function, then shuffle
their marks randomly and recalculate the marked correlation function
100 times. For each radial bin, we calculate the mean of these 100
realizations of the marked correlation function and the standard
deviation between realizations. The mean plus or minus one standard
deviation is given by the dotted lines.

We have also tried to quantify the systematic error induced by
including haloes over such a large mass range. We repeat the procedure
used to obtain the dotted lines of Fig.~\ref{fig:onehilomass}a, but
instead of shuffling marks over our entire sample of haloes, we sort
the haloes into eight mass bins. Then we only shuffle the marks within
each bin in mass. Therefore, although a halo receives the mark of a
random halo in the sample, it is only permitted to receive the mark of
a halo with a very similar mass. We then take the mean and standard
deviation in each radial bin of the realizations of the marked
correlation function as before. This binning procedure makes very little
difference, in fact, and gives us confidence that the
$\tilde{\omega}_\mathrm{f}$ parameter scales out the mass dependence
of halo formation redshifts sufficiently well for the purposes of this
test.

To give a numerical indication of the strength of the signal, we
calculated the marked correlation function in one large bin between
$1$ and $5\ h^{-1}\ \mathrm{Mpc}$ and estimated the error using the
same shuffling procedure as before. This indicated that the value of
$\xi_\mathrm{marked}$ was inconsistent with unity at the $5\sigma$
level. It is the large volume of our simulation which enables us to
see a signal in the marked correlation function of such massive
haloes, but we find that the behaviour of samples of haloes of lower
mass is similar. Moreover, the dynamic range of the simulation is such
that we can study relatively small haloes, robustly determining
formation times of haloes down to a mass of \hbox{$5.5\times 10^{10}\
h^{-1}\ \mathrm{M}_\odot$}. For galaxy-sized haloes with between 500
and 2000 particles, for example, we see a larger environmental
dependence. The marked correlation function for haloes of this mass is
given in Fig.~\ref{fig:onehilomass}b. The abundance of haloes of this
mass means that the error in the determination of the marked
correlation function is negligible at most scales of interest. The
excess at small separations is more significant than for the more
massive haloes, and the size of the effect is also larger. This is
qualitatively consistent with \citet{GAO05} since the effect for which
they tested (a variation in clustering amplitude with halo formation
redshift) was larger for haloes of lower mass.

\begin{figure}
  \begin{center}
    \leavevmode
    \psfig{file=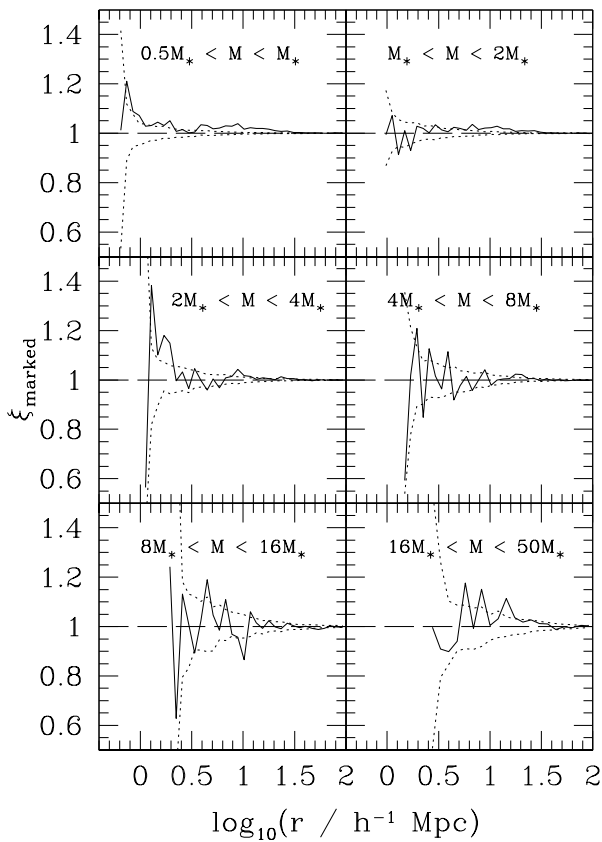,width=8cm}
    \caption{The marked correlation function of haloes with mass in
    the ranges shown (solid lines), with $\tilde{\omega}_\mathrm{f}$
    as the mark. Dotted lines are calculated as in
    Fig.~\ref{fig:onehilomass}a. The top four panels may be compared
    with fig.~4 of \citet{SHE04b} since the value of $M_\ast$ in their
    simulation is approximately twice ours. Note, though, the
    difference in axis scale, and the fact that in their figure
    formation redshift is used for the mark, whereas here
    $\tilde{\omega}_\mathrm{f}$ is used (see Fig.~\ref{fig:diffmarks}
    which shows this does not affect our conclusions). The lower four
    panels cover the same mass range as
    Fig.~\ref{fig:onehilomass}a.}\label{fig:sixpanel}
  \end{center}
\end{figure}

Splitting the mass range used in Fig.~\ref{fig:onehilomass}a into four
parts gives the result shown in the lower four panels of
Fig.~\ref{fig:sixpanel}. Firstly, it is clear that the estimates of
\hbox{$\xi_\mathrm{marked}$} in Fig.~\ref{fig:sixpanel} are far more
noisy; while the mass range covered in Fig.~\ref{fig:onehilomass}a
contains 34\,241 haloes in the Millennium Simulation, the lower four
panels of Fig.~\ref{fig:sixpanel} cover mass ranges containing, in
order of increasing mass, 18\,384, 9172, 4298 and 2387 haloes
respectively. Since the quality of the statistics is governed by the
number of halo \emph{pairs}, the effect is noticeable even given the
large volume of the simulation. This highlights the importance of
properly scaling out the mass dependence of halo formation redshift,
so that we may average over large ranges in halo mass.

For similar reasons, (i.e.\ the effect of cross-correlations between
bins) the marked correlation function for the whole mass range of
Fig.~\ref{fig:onehilomass}a is not simply the average of the marked
correlation function of each of the four sub-ranges.  For example, if
we perform the test described above of calculating the marked
correlation function for one large bin between $1$ and $5\ h^{-1}\
\mathrm{Mpc}$, we see that the function for the range
$2M_\ast<M<4M_\ast$ is greater than unity only at the 1$\sigma$ level.
In the highest mass range, the function in this radial bin is less
than unity, by approximately $1.5\sigma$.  One would normally dismiss
this apparent change in the sign of the environmental dependence as
insignificant, especially given our free choice of bin size and the
freedom in the definition of the halo catalogue and merger trees, but
it is qualitatively consistent with fig.~4 of \citet{SHE04b}. In the
amalgamated sample, of course, most of the halo pairs which include a
member in the highest mass bin have one member of the pair from a
lower mass bin. It may therefore still be the case that the product of
the marks of the haloes in such a pair with separation $r$ is usually
greater than $\bar{m}^2$, and yet a halo pair of separation $r$ in
which both members are from the highest mass bin usually gives a
product of marks less than $\bar{m}^2$. This is a barrier to the clean
interpretation of these results, since when measuring the
environmental dependence of haloes in some mass range, the environment
can only be defined in terms of haloes in the same mass range. We
address this problem by explicitly separating the `tracer' population
from the `marked' population in Section~\ref{subsec:cross} below.

Recall that the small-scale cutoff in the marked correlation function
occurs because there are no haloes in the given mass range which occur
at such small separations in the simulation: an exclusion effect. The
radius at which this occurs depends on mass, and certainly this effect
is noticeable when comparing the top-left panel of
Fig.~\ref{fig:sixpanel} to the bottom-right panel.

This dependence on halo mass of the scale upon which we can measure
environment again suggests separating the tracer and marked
populations, as we do when calculating a marked cross-correlation
function below. Of course, some dependence is inevitable since more
massive haloes tend to have larger radii. This reinforces the point
that a method in which we choose beforehand a fixed scale on which to
measure environment -- looking at scales at which there is a peak in
the marked correlation function for low mass haloes, say -- may be
flawed, since the outer regions of more massive haloes will contribute
to the definition of their own environment.

\begin{figure}
  \begin{center}
    \leavevmode
    \psfig{file=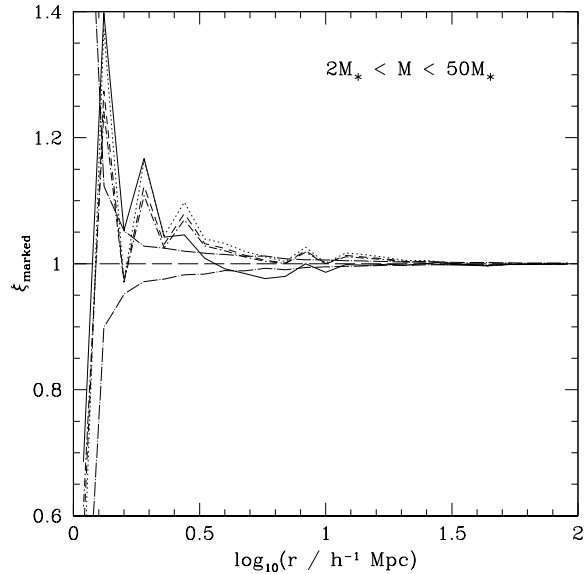,width=8cm}
    \caption{The marked correlation function with several different
    choices of mark. In each case the population of haloes used is the
    same as in Fig.~\ref{fig:onehilomass}a, but we vary the choice of mark
    as follows: solid line -- formation redshift; short-dashed line --
    $\tilde{\omega}_\mathrm{f}$; dot-dashed line -- formation redshift
    divided by the mean formation redshift for haloes of that mass
    (determined from the simulation); dotted line -- haloes are ranked
    by $\tilde{\omega}_\mathrm{f}$ then reassigned a mark (preserving
    this ranking) such that the marks follow the analytic distribution
    for $\tilde{\omega}_\mathrm{f}$ given by the thick, solid line in
    Fig.~\ref{fig:dpdzw}b. The long-dashed line through
    \hbox{$\xi_\mathrm{marked}=1$} is shown to guide the eye. The dispersion
    in $\xi_\mathrm{marked}$ in the scrambled catalogues is shown only
    for a mark of $\tilde{\omega}_\mathrm{f}$ (dot-long-dashed lines),
    since it is very similar in each case.}\label{fig:diffmarks}
  \end{center}
\end{figure}

We emphasized earlier the importance of being able to calculate a
marked correlation function for a sample of haloes which spans a large
range in mass, and suggested a scheme for scaling out the mass
dependence of halo formation times based on the analytic work of
\citet{LAC93}. One can easily imagine other ways to scale out this
dependence, however, and we attempt to show the difference between
various methods in Fig.~\ref{fig:diffmarks}. For the solid line we
make no attempt to correct for the mass dependence of halo formation
redshift and simply use $z_\mathrm{f}$ as the mark, while for the
other three lines some kind of scaling is applied. The short-dashed
line is the result for our fiducial mark,
$\tilde{\omega}_\mathrm{f}$. The dot-dashed line uses the simulation
itself to determine the scaling: we simply divide the formation
redshift for each halo by the mean formation redshift for haloes of
that mass. This does not take into account changes in the shape of the
distribution of formation redshift as a function of mass. For the
dotted line we first calculate $\tilde{\omega}_\mathrm{f}$ for each
halo, as above. Then we rank the haloes in order of
$\tilde{\omega}_\mathrm{f}$ and reassign each one a mark, preserving
the ranking, such that the final distribution of marks is precisely
the analytic distribution given by the thick, solid line of
Fig.~\ref{fig:dpdzw}b. This explicitly enforces near-mass
independence, hopefully without distorting the shape of
$\xi_\mathrm{marked}$ too much since the shape of the analytic
distribution matches the measured distribution quite well. It seems
from Fig.~\ref{fig:diffmarks} that any reasonable method for scaling
out the mass dependence of the distribution of halo formation redshift
gives similar results.

The errors in the marked correlation functions measured with these
four different marks are very similar, and the effect of shuffling
only within narrow mass bins remains small in each case. Indeed, when
we force the marks to follow the analytic distribution for
$\tilde{\omega}_\mathrm{f}$ we might expect it to make no difference
whether we shuffle between haloes of all masses or only between haloes
of similar mass, and we have checked that this is indeed the
case. Fig.~\ref{fig:diffmarks} gives us confidence that our
conclusions about the environmental dependence of halo formation
redshift are robust to changes in the precise definition of the mark,
so long as the mark remains a reasonable proxy for the halo formation
redshift as defined in Section~\ref{subsec:mtrees}, and so long as the
width of the distribution of marks remains similar. We conclude,
therefore, that we have significant evidence that halo formation
redshift does depend on environment, and we explore this in more
detail in what follows.

\subsection{A marked cross-correlation function}\label{subsec:cross}

Even the marked correlation functions we calculate above which include
haloes in a wide range of mass (up to a factor of about 25 between the
lowest and highest mass) utilise only a fraction of the dynamic range
available in the Millennium Simulation. We are more limited by the
fact that the wider the range of mass studied, the harder the marked
correlation functions are to interpret. If we include very small
haloes, then because low mass haloes are more abundant, the function
will be dominated at all scales by contributions from low mass
haloes. The contribution from haloes of any given mass only cuts in
above some scale determined by the exclusion effect from the non-zero
size of the halo. On the other hand, if we wish to study the
environmental dependence of the formation times of only very massive
haloes, we will have poor statistics even when simulating enormous
volumes, and it will not be clear in any case that such massive haloes
are good tracers of environment. We have attempted to address some of
these problems by defining a marked cross-correlation function.

Consider two populations of haloes, which we denote the `tracer'
population and the `marked' population. We then define the marked
cross-correlation function,
\hbox{$\xi_\mathrm{marked}^\mathrm{cross}(r)$}, by
\begin{equation}
\xi_\mathrm{marked}^\mathrm{cross}(r)=
\sum_{\{i,j\ |\ r_{ij}=r\}}\frac{m_j}{n(r)\bar{m}}\quad ,
\end{equation}
where the sum is now taken over pairs \hbox{$\{i,j\}$} such that halo $i$ is
from the tracer population and halo $j$ is from the marked population,
\hbox{$n(r)$} is the number of such pairs of separation $r$ and $\bar{m}$ is
the mean mark of the haloes in the marked population. This tells us
about the environmental dependence of the mark in the marked
population, with environment defined in terms of the tracer
population. It retains the property that a deviation of the function
from unity indicates environmental dependence. Note, however, that it does
not have some of the properties of a normal cross-correlation
function: it will be different if we exchange the marked and tracer
populations, and the marked cross-correlation function of a population
with itself is not equivalent to the marked autocorrelation function.

\begin{figure}
  \begin{center}
    \leavevmode \psfig{file=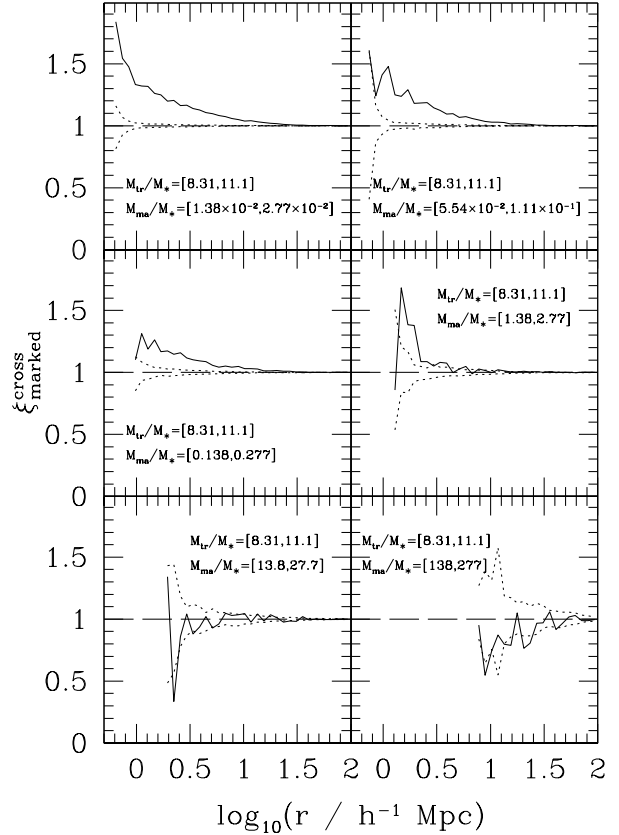,width=8cm}
    \caption{Solid lines show the marked cross-correlation function of
    haloes with the same tracer population (\hbox{$6\times 10^{13}\
    h^{-1}\ \mathrm{M}_\odot$} haloes) each time, but with a marked
    population of different mass in each panel. The mark is
    $\tilde{\omega}_\mathrm{f}$. The haloes in the tracer population
    have mass $M_\mathrm{tr}$ in the range shown, while those in the
    marked population have mass $M_\mathrm{ma}$ in the range shown
    (recall an $M_\ast$ halo contains 7221 particles, and a
    galaxy-sized, 1000 particle halo has mass \hbox{$0.138
    M_\ast$}). We again show the dispersion in 100 calculations of the
    marked cross-correlation function (dotted lines), shuffling the
    marks at random between haloes each time.}\label{fig:sixcross}
  \end{center}
\end{figure}

Fig.~\ref{fig:sixcross} gives six examples of marked cross-correlation
functions with $\tilde{\omega}_\mathrm{f}$ as the mark. We estimate
the dispersion among realizations of the functions by recalculating
the function 100 times with the marks shuffled, as before. The tracer
population is the same in each panel, but the mass of the marked
population increases from left to right and from top to bottom. For
the higher mass populations there seems to be a trend that as the mass
of the marked population increases, the positive signal from the
marked cross-correlation function becomes weaker, perhaps even
changing sign when the mass of the marked haloes becomes greater than
that of the tracer haloes. Since we expect individual $L_\ast$
galaxies to occupy haloes containing approximately 1000 particles in
this simulation (a halo with 1000 particles has a mass of \hbox{$0.138
M_\ast=8.61\times 10^{11}\ h^{-1}\ \mathrm{M}_\odot$}), the results
for lower mass populations suggest we have significant evidence that
galaxy-sized haloes near \hbox{$6\times 10^{13}\ h^{-1}\
\mathrm{M}_\odot$} haloes have earlier formation times than the mean.

\begin{figure}
  \begin{center}
    \leavevmode
    \psfig{file=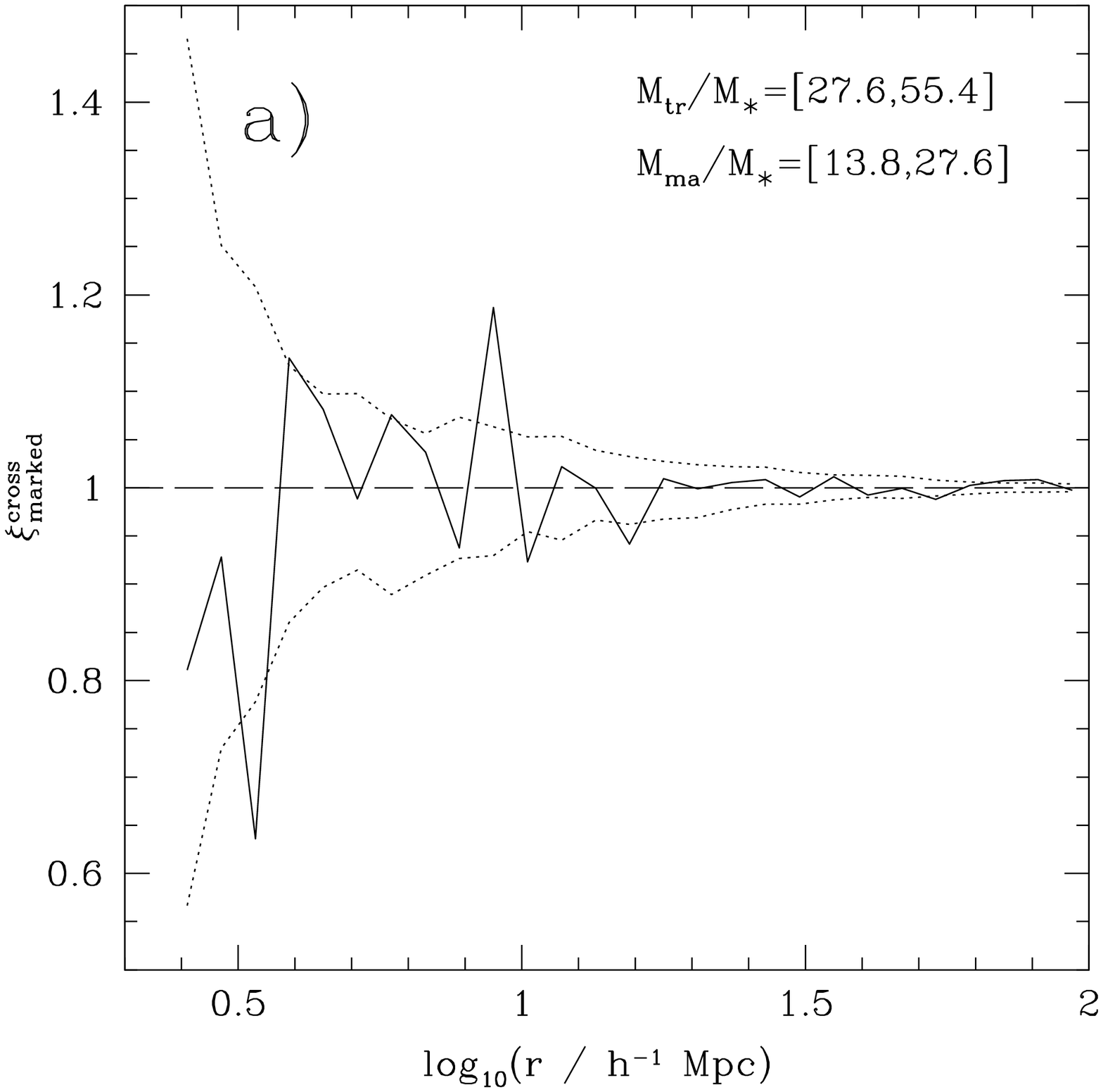,width=8cm}
    \psfig{file=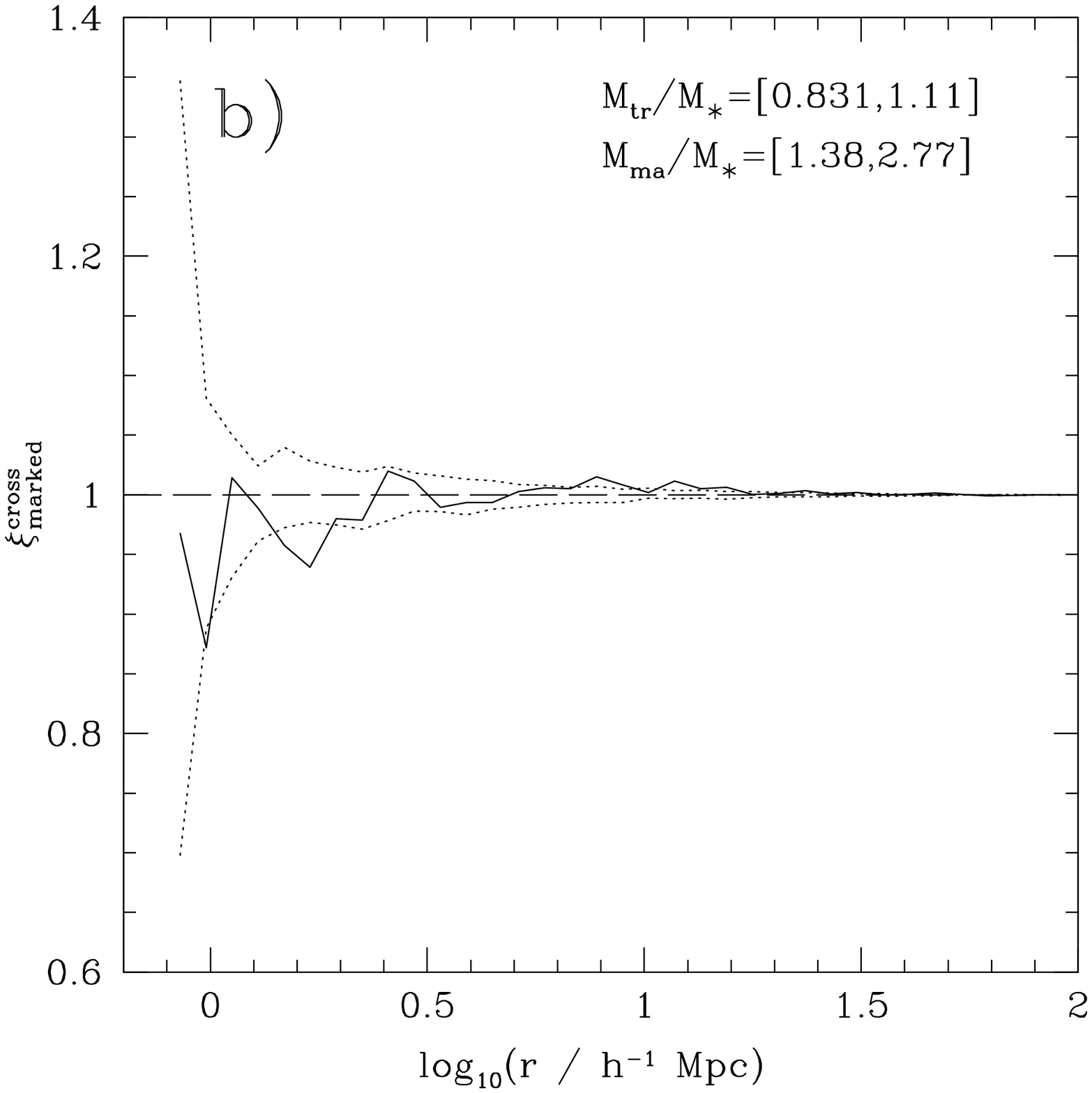,width=8cm}
    \caption{The top panel shows the marked cross-correlation function
    using the same marked population as the lower-left panel of
    Fig.~\ref{fig:sixcross}. Here, though, we choose the tracer
    population to be more massive than in Fig.~\ref{fig:sixcross},
    and, importantly, more massive than the marked population. The
    dotted lines are as in Fig.~\ref{fig:sixcross}. The bottom panel
    uses the same marked population as the middle-right panel of
    Fig.~\ref{fig:sixcross} but with a lighter tracer
    population.}\label{fig:hltrace}
  \end{center}
\end{figure}

Comparing to the marked autocorrelation function, then, the most
puzzling panels of Fig.~\ref{fig:sixcross} are the lower right panel
and, to a lesser extent, the lower left panel. The trend in the marked
cross-correlation function in the lower right panel is in the opposite
sense to that which one may expect having seen the earlier results
(less than unity at $\sim\! 2.5\sigma$ for one bin between $5$ and
$30\ h^{-1}\ \mathrm{Mpc}$), while we see no signal in the lower left
panel. Recall that we expect more close pairs in more dense
regions. This means the marked correlation function at small scales is
representative of dense regions, so that we expect to see earlier
formation times. This expectation may not hold in the situation
represented by the lower right panel of Fig.~\ref{fig:sixcross},
however. While the tracer population consists of haloes with mass near
\hbox{$6\times 10^{13}\ h^{-1}\ \mathrm{M}_\odot$,} the marked
population in this panel consists of very massive haloes, of around
\hbox{$10^{15}\ h^{-1}\ \mathrm{M}_\odot$}. These large haloes will be
found only in regions which are at least moderately dense, and many
will be found in the very densest parts of the simulation: in the core
of the filaments making up the cosmic web, or at the intersection of
the filaments. In these highly dense regions, we expect nearby haloes
to also be very massive, whereas it is in the moderately dense regions
that $6\times 10^{13}\ h^{-1}\ \mathrm{M}_\odot$ haloes are most
abundant. It may be that by choosing this tracer population, the close
pair counts are dominated by haloes in only moderately dense regions,
since it is here that our tracer population is most abundant. The
large-scale pair counts are more representative of the average
environment of \hbox{$10^{15}\ h^{-1}\ \mathrm{M}_\odot$} haloes,
which is even more dense. If this interpretation is correct, we might
anticipate that using a more massive tracer population would reverse
the trend, so that $\xi_\mathrm{marked}^\mathrm{cross}(r)$ was once
again larger on small scales. We test this prediction in
Fig.~\ref{fig:hltrace}a. In this figure the marked population is the
same as in the lower-left panel of Fig.~\ref{fig:sixcross}, since this
allows us to choose a sufficiently abundant tracer population that is
nevertheless more massive than the marked population.

While the signal we see in Fig.~\ref{fig:hltrace}a is weak and more
noisy (with only 872 haloes in the tracer population and 2189 haloes
in the marked population) there is no repetition of the unexpected
trend seen in the lower panels of Fig.~\ref{fig:sixcross}. We have
also performed the converse test, in Fig.~\ref{fig:hltrace}b. That is,
we take the marked population that produces a positive signal in the
middle right panel of Fig.~\ref{fig:sixcross}, and find the marked
cross-correlation function of these haloes with a less massive tracer
population. Using lower mass haloes also improves our statistics:
there are $26\, 417$ and $20\, 968$ haloes in the tracer and marked
population respectively. The positive signal seen in
Fig.~\ref{fig:sixcross} at small scales is wiped out, and if anything
there is a weak negative signal. This suggests that a definition of
environment using some tracer population only really corresponds with
our intuition of what environment should mean (close pairs
representing a dense environment) if the tracer population is at least
as strongly clustered as the marked population.

\subsection{A simpler test of environment}\label{subsec:oldtest}

\begin{figure}
  \begin{center}
    \leavevmode
    \psfig{file=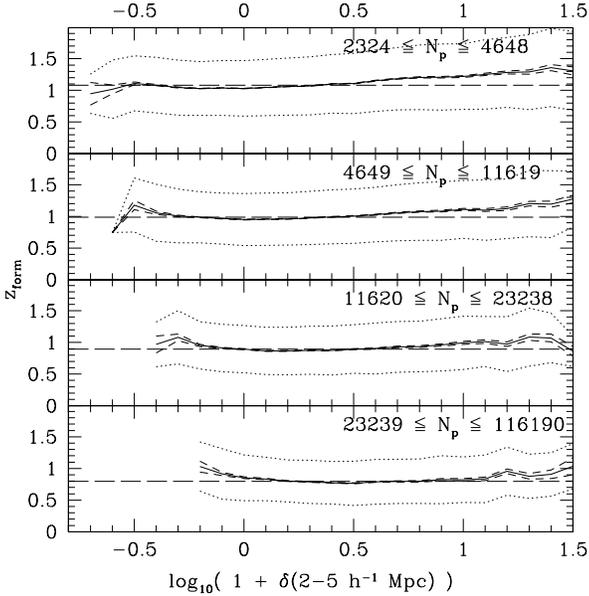,width=8cm}
    \caption{The formation redshift of haloes as a function of
    overdensity in a spherical shell of inner radius $2\ h^{-1}\
    \mathrm{Mpc}$ and outer radius $5\ h^{-1}\ \mathrm{Mpc}$ centred
    on the halo (solid lines). The range of particle numbers for the
    haloes in each panel is shown; these are chosen so that haloes
    have the same mass as those in the corresponding panel of fig.~3
    of \citet{LEM99}. Note the difference in the scale of the
    horizontal axis between the linear scale of fig.~3 of
    \citet{LEM99} and the logarithmic scale of this figure which
    extends to higher densities. Short-dashed lines show the error on
    the determination of the mean formation redshift in each bin in
    overdensity.  Dotted lines show the 1-$\sigma$ dispersion in halo
    formation times.  The flat, long-dashed line is at the mean
    formation redshift for all haloes in that bin in mass, and is
    shown only to guide the eye.}\label{fig:lkfig}
  \end{center}
\end{figure}

Having seen evidence of environmental dependence of halo formation
times in the marked correlation function, it is interesting to see
whether the volume and dynamic range offered by the Millennium
Simulation allow us to see a signal in other measures of environment.
For Fig.~\ref{fig:lkfig} our measure is simply the overdensity in dark
matter in a spherical shell between $2$ and \hbox{$5\ h^{-1}\
\mathrm{Mpc}$} from the centre of the halo (where the centre is
defined, as before, as the position of the particle in the main
substructure of the halo having the least gravitational potential
energy). This is the same measure as used in fig.~3 of \citet{LEM99}
in which no signal is apparent, despite the simulation being the same
as the one which showed evidence of environmental dependence in the
marked correlation function analysis of \citet{SHE04b}: both studies
used the GIF simulations \citep{JEN98,KAU99}. The range in halo mass
used for each panel of our plot is the same as in fig.~3 of
\citet{LEM99}. A clear trend is visible; for the top three panels
especially, there is evidence that haloes in regions with
overdensities greater than about 1 or 2 have higher formation
redshifts. We can follow this trend over a very wide range in
overdensity.

\begin{figure}
  \begin{center}
    \leavevmode
    \psfig{file=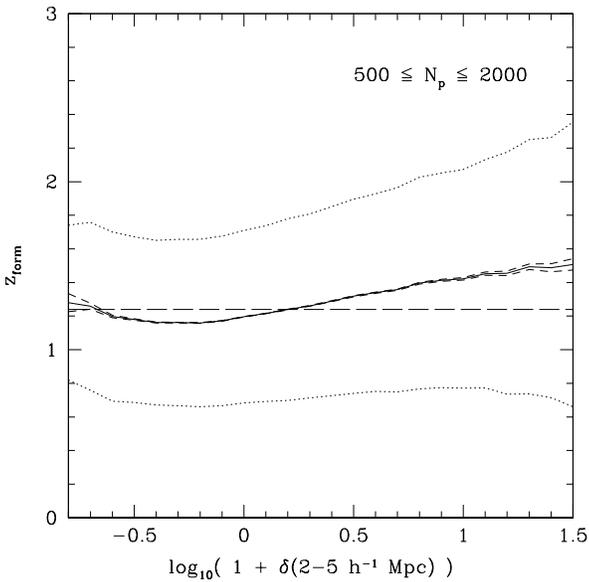,width=8cm}
    \caption{As Fig.~\ref{fig:lkfig}, but for haloes with mass such
    that we may expect them to host a single bright galaxy. The halo
    sample is the same as for
    Fig.~\ref{fig:onehilomass}b.}\label{fig:lklowm}
  \end{center}
\end{figure}

Because of the high resolution of our simulation, we may extend this
technique to lower mass haloes. Haloes which are expected to host a
single, bright galaxy (and -- importantly for this analysis -- the
progenitors of these haloes) are well resolved, containing roughly
1000 particles. Fig.~\ref{fig:lklowm} is similar to a single panel of
Fig.~\ref{fig:lkfig}, but using haloes with between 500 and 2000
particles, corresponding to masses of between $4.30\times 10^{11}$ and
\hbox{$1.72\times 10^{12}\ h^{-1}\ \mathrm{M}_\odot$}. It is clear we
have very significant evidence that haloes in denser regions have
higher formation redshifts than the mean, and conversely that haloes
in less dense regions have lower formation redshifts than the
mean. The size of the effect is similar to that for the more massive
haloes (larger, if anything -- consistent with \citet{GAO05}), but is
detected more cleanly due to the large sample size. Reproducing
Figs~\ref{fig:lkfig} and~\ref{fig:lklowm} using
$\tilde{\omega}_\mathrm{f}$ as a proxy for formation redshift gives
extremely similar results. The mean is slightly offset, as one would
expect from Fig.~\ref{fig:zvsnp}, but the trends are identical.

The dispersion in formation times at a given overdensity is larger
than the systematic variation between different
overdensities. Therefore it is unclear from these data what the effect
of this variation will be on, for example, the properties of the
central galaxies hosted by these haloes. \citet{GAO05} use the
Millennium Simulation to address the question of how the clustering of
haloes of a given mass depends on their formation time, and find a
clear difference between the clustering of the oldest and youngest
haloes. We do not yet know the effect of this difference on
observables such as the galaxy correlation function (split by colour
or galaxy environment), especially in the light of the fact that
galaxies of a given luminosity reside in haloes with a range of mass,
with most galaxy light expected to come not from single-occupation
haloes such as these, but from group-sized haloes \citep{EKE05}. This
problem may be addressed by galaxy catalogues constructed using the
halo catalogues and merger trees from the Millennium Simulation
\citep[e.g.,][]{BOW06,CRO06}.

In Fig.~\ref{fig:lklowm}, the mean formation redshift crosses from
being being below the global mean for haloes in this mass range
(long-dashed line) to above it at an overdensity \hbox{$\delta\approx
0.6$}. It appears this crossing point moves to higher overdensity for
more massive haloes: in the bottom panel of Fig.~\ref{fig:lkfig} the
crossing point is at \hbox{$\delta\approx 5$}. We have repeated these
calculations and our marked correlation function calculations using
different halo catalogues and merger trees, including those of
\citet{GAO05}. While the position of this crossing point and the
precise shape of the curve are sensitive to the detailed definition of
the halo catalogues and the merger trees, the general trends are
robust.

Performing the calculations of \citet{GAO05} using our trees, or our
marked correlation function analysis using their trees, gives
qualitatively consistent results. This is encouraging since the two
sets of trees are constructed quite differently, though using the same
{\small SUBFIND} catalogue. For example, they define the formation
time using the mass within $r_{200}$ (the radius at which the enclosed
density falls below 200 times the critical density), whereas we use
the mass of the friends-of-friends halo. Also, to find the merger tree
of a friends-of-friends halo they follow only the merger history of
its main substructure, while we follow the combined histories of each
of the substructures which make up the halo. This allows \citet{GAO05}
to track more easily the history of a halo which was temporarily the
substructure of a larger halo, but which has since de-merged to become
a separate halo in its own right. It is important to deal with these
de-mergers well, since they lead to close pairs in dense environments
in which one member of the pair is likely to be unusual in some way:
for example, it may be assigned an artificially low formation
redshift. We note in Section~\ref{subsec:mtrees} that we take
precisely one merger tree per friends-of-friends halo, discarding the
lower mass trees which were split off having been deemed to have been
spuriously connected. Including these trees causes de-merger problems
since merger tree haloes can split despite remaining in the same
friends-of-friends halo. On the other hand, using merger trees
constructed purely from friends-of-friends catalogues without
identifying substructure causes even greater problems, since haloes
are often spuriously attached and subsequently split. So long as we
take these de-mergers into account, all our results remain robust to
the precise definition of the halo catalogue or the merger trees.

\section{Conclusions}\label{sec:conc}

In this paper, we have looked for evidence of an environmental
dependence of halo formation times, using what we consider to be an
especially sensitive test, and using a very large simulation which
offers excellent statistical power when constraining the properties of
haloes with a large range of mass. We have very strong evidence that
haloes of a given mass in denser regions formed at higher redshift
than those in less dense regions. This result is robust to changes in
the mark used as a proxy for formation redshift, and we conclude that
the observed dependence is not affected by systematic bias from
averaging over a range of halo mass. Our conclusions are also
unaffected by the precise definition of the halo catalogue or by the
details of the construction of the merger trees.

Separating the haloes for which we wish to measure environment from
those used to define environment allows us to look for the origin of
the signal in more detail. We see a stronger dependence on
environment for low mass haloes, although the effect is still present
when more massive haloes are considered. We also note that in this
context it only makes sense if the environment of low mass haloes is
traced by a population of higher mass haloes. Using numerous, low mass
haloes to trace the environment of more massive haloes means that our
definition of environment may no longer correspond to an intuitive
definition, in that it may no longer be the case that a relatively
large number of close neighbours implies a relatively dense
environment.

If we revert to a more intuitive test of the dependence of formation
time on environment, and look at the mean formation redshift of haloes
of a given mass as a function of the local overdensity in dark matter,
we note that the size and resolution of our simulation allows us to
see a highly significant signal of environmental dependence for haloes
with a wide range in mass, but again especially for low mass
haloes. We are able to perform this test for haloes which we expect to
host only a single, bright galaxy, since the progenitors of these
haloes are well resolved. The size of the variation in mean formation
redshift is smaller than the (large) dispersion in formation redshift
for haloes residing in a region of given overdensity. This makes the
impact of this dependence on statistics such as the galaxy correlation
function unclear, though this effect is studied in more detail by
\citet{GAO05} (who used the same simulation but different merger
trees) where the age dependence of halo clustering is studied and a
significant signal is observed.

Our results have, in any case, some implications for galaxy formation
models and for halo models of clustering. Any simple version of the
halo occupation distribution formalism \citep{SEL00,BER02,COO02}, for
example, has as one of its basic assumptions that knowing the mass of
a halo is sufficient to statistically determine the properties of its
galaxy population. So long as the properties of the galaxy population
depend sufficiently strongly on the merger history of a halo, we see
that this assumption is no longer strictly valid, and this therefore
calls into question the validity of results based on this formalism
\citep[e.g.,][]{BER03,ABA05}. While extended Press-Schechter theory
does a reasonable job of predicting the distribution of halo formation
redshifts when averaging over haloes in all environments, it also
predicts that the formation history is independent of environment. We
clearly see that this is not the case, so the practice of assigning a
Monte Carlo merger tree constructed according to extended
Press-Schechter theory to a simulated halo based only on the halo mass
is called into question. The magnitude of this effect on any
observables drawn from mock galaxy catalogues generated by
semi-analytic models using these merger trees is unclear at this
stage, and may only become clear when comparing catalogues produced
using Monte Carlo merger trees with those produced using merger trees
extracted directly from the simulation being populated. This latter
approach has become feasible with the advent of simulations of
sufficient resolution and volume, such as the Millennium Simulation
used here. It may still be the case that the width of the distribution
of formation redshifts in a given environment, and the scatter in
other relations such as the halo mass -- central galaxy luminosity
relation, wash out this effect. Uncertainties in the galaxy formation
models themselves may prove to be more important. Equally, though, if
other halo properties such as the concentration and angular momentum
depend strongly enough on formation time or environment, then this may
help the models to better match and explain observations of the
environmental dependence of galaxy colour and morphology, or the
concentration or velocity profiles of galaxies of different ages.

\section*{Acknowledgements}

GH acknowledges receipt of a PhD studentship from the Particle Physics
and Astronomy Research Council. The simulation used in this paper was
carried out as part of the programme of the Virgo Consortium on the
Regatta supercomputer of the Computing Centre of the
Max-Planck-Society in Garching.

\bibliography{allbib}
\end{document}